\documentclass[final,3p,times,twocolumn]{elsarticle}

\usepackage{graphicx}

\usepackage{amssymb}

\usepackage{lineno}

\newcommand{\rmi}{\mathrm{i}}

\journal{Physica C}

\begin{document}

\begin{frontmatter}



\title{Vortex motion in Nb/PdNi/Nb trilayers: new aspects in the flux flow state}


%
\author[RomaTre]{K. Torokhtii}
\author[Salerno]{C. Attanasio}
\author[Salerno]{C. Cirillo}
\author[Salerno]{E. A. Ilyina}
\author[RomaTre]{N. Pompeo}
\author[Sapienza]{S. Sarti}
\author[RomaTre]{E. Silva\corref{cor}}
\cortext[cor]{Corresponding author.}
\ead{silva@fis.uniroma3.it}
\address[RomaTre]{Dipartimento di Fisica ``E. Amaldi'' and Unit\`a CNISM,
Universit\`a Roma Tre, Via della Vasca Navale 84, 00146 Roma,
Italy}

\address[Salerno]{CNR-SPIN Salerno and Dipartimento di Fisica ``E. R. Caianiello'', Universit\`a degli Studi di Salerno, I-84084 Fisciano, Salerno, Italy}

\address[Sapienza]{Dipartimento di Fisica, Universit\`a "La Sapienza", 00185 Roma, Italy}

\begin{abstract}

We study the dynamics of vortex lines in Supercondutor/Ferromagnet/Superconductor (SFS) heterostructures at microwave frequencies. We have employed swept-frequency, Corbino-disk and resonant, dielectric-resonator techniques to obtain the field and temperature dependence of the vortex-state parameters. We concentrate here on the genuine flux-flow resistivity $\rho_{ff}$, that we access at subcritical currents using a sufficiently high driving frequency. We find that  $\rho_{ff}$ does not follow the well-known Bardeen-Stephen model. Instead, it is well described by a full time-dependent Ginzburg-Landau expression at very thin F layer thickness, but changes to a previously unreported field-dependence when the F layer exceeds a few nm. 

\end{abstract}

\end{frontmatter}


\section{Introduction}
\label{intro}
The dynamic properties of the vortex state have been studied extensively with experiments, theory and numerical simulations, with renewed interest after the discovery of cuprate superconductors. Complex vortex matters have been identified, and their properties theoretically investigated \cite{blatterRMP94}. In face of the enormous amount of progress in the field, the simplest, free flux flow (FFF) has received much less attention, despite its direct connection to the physics in the vortex core. The discovery of Iron-based superconductors, and the revamped interest in superconductor/ferromagnet hybrid structures make this issue even more interesting than before.

Experiments on the vortex motion resistivity often compare the flux-flow state to the prediction of the Bardeen-Stephen (BS) model, which predicts
\begin{equation}
\label{eq:rhoBS}
    \rho_{ff,BS}=c \rho_n {B}/{\mu_0 H_{c2}}
\end{equation}
where $\rho_n$ is the normal state resistivity, $B$ is the field induction, $H_{c2}$ is the temperature-dependent upper critical field and $c\sim 1$. This behavior is often observed, apart the field region approaching $H_{c2}$, where additional effects, e.g. the suppression of the condensate, increase $\rho_{ff}$ above $\rho_n B/\mu_0H_{c2}$. 

In this paper we study the flux flow resistivity $\rho_{ff}$ in SFS trilayers using high-frequencies currents, of the order of several GHz. The motivation resides in the possibility to access the genuine flux-flow state (defined as irrelevance of pinning and free motion of vortices, hindered solely by the viscous force) without the need for large currents. In fact, motion of vortices is a complex matter that involves interaction with defects, relation with the vortex lattice elastic moduli, and the nature of the vortex core (responsible for the viscous drag). However, high frequency measurements prove useful in simplifying the physical problem: at several GHz vortices oscillate over distances that are typically much shorter than the vortex lattice constant and of the defect average distance \cite{tomaschPRB88}. Thus, they remain mostly near their equilibrium position. In most cases one can apply safely the mean-field theory developed for vortex dynamics  \cite{GR,CC,brandtPRL91,MStheory}. Interestingly, a single formulation is able to describe many of the mentioned theories. In fact, the vortex motion resistivity can be described by the following expression \cite{pompeoPRB08}:
\begin{equation}
\label{eq:rhovm}
    \rho_{vm}=\rho_{vm,1}+\rmi\rho_{vm,2}=\rho_{ff}\frac{\varepsilon+\rmi\left(\nu/\bar{\nu}\right)}{1+\rmi\left(\nu/\bar{\nu}\right)}
\end{equation}
\noindent where the resistivity $\rho_{ff}$ represents the free flux flow value, reached in the high frequency limit in which vortices experiences only the dissipative viscous drag, $\bar{\nu}$ is a characteristic frequency and the dimensionless parameter $0\leq\varepsilon\leq 1$ is a measure of thermal activation phenomena. For the purposes of this paper, we explicitly note that, when the measuring frequency $\nu\gg\bar{\nu}$, $\rho_{vm}\simeq \rho_{ff}$, irrespective of pinning and creep (whose relevance are within $\bar{\nu}$ and $\varepsilon$). Since in most superconductors the characteristic frequencies lay at or below the microwave frequencies, an experimental investigation at several GHz is a powerful tool to access the genuine $\rho_{ff}$.
\section{Samples and experimental setup}
\label{Setup}
\noindent
Nb/Pd$_{84}$Ni$_{16}$/Nb trilayers have been grown by UHV dc magnetron sputtering at room temperature, as reported elsewhere \cite{cirilloSUST11}. Samples were grown onto sapphire substrates to avoid the deleterious contribution of semiconducting substrates to the microwave measurements \cite{pompeoSUST05}. The nominal thickness, calibrated with the deposition rate, was as follows: all trilayers had Nb thickness $d_{Nb}$=15 nm for both upper and lower layers. In this work we concentrate on the samples with PdNi thickness $d_F$= 1, 2, 9 nm. dc resistivity measurements in samples grown on the same conditions consistently yielded a constant normal state $\rho_{n}\simeq 22\pm 5\; \mu\Omega$cm for temperatures $T_c<T<30\;$K. Critical temperatures of 6.25 K, 5.1 K and 4.25 K for the samples with $d_F$= 1, 2, 9 nm, respectively,  were evaluated by the disappearance of the microwave signal and were consistent with dc measurements with a criterion of 90\% of the normal state.

We have used two different setups in order to measure the microwave response of SFS heterostructures. The Corbino disk setup was used to obtain $\rho_{ff}$ from the frequency dependence of $\rho_{vm}$ between 2 and 20 GHz, in SFS structures with $d_F=$1 and 2 nm. The setup and the measuring technique has been extensively described previously \cite{silvaSUST11}, and we report here only the results. In this context, it is important to stress that the swept-frequency data directly yielded a characteristic frequency $\bar{\nu}$ which decreased with $d_F$, and was as low as 4 GHz in the sample with $d_F=$2 nm. Thus, we have chosen to perform measurements in  trilayers with $d_F=2$ and 9 nm with a more sensitive dielectric resonator at the operating frequency $\nu_0 \simeq$ 8 GHz.
\begin{figure}[h]
\centerline{\includegraphics[width=8.5cm]{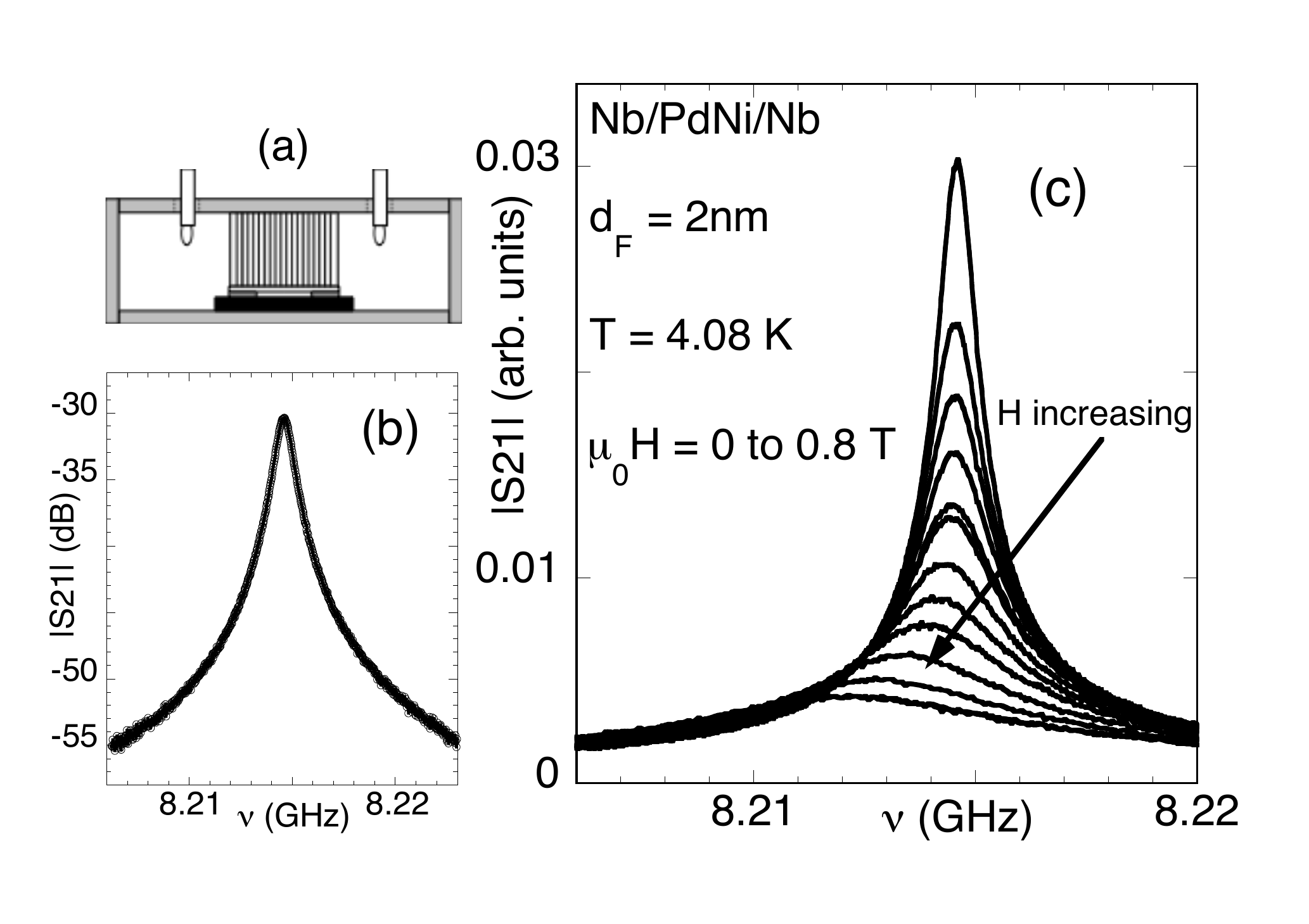}}
\vspace{-5mm}
  \caption{(a) sketch of the section of the DR resonator (proportions are approximate). Rutile rod: hatched area; sample: black rectangle. (b) log plot of the resonance curve for the trilayer with $d_F=$2 nm at $T=$4.08 K and in zero field. (c) field-variation of the resonance curve.}
\label{figresonator}
\end{figure}

The dielectric resonator (DR) is based upon a rutile (TiO$_2$) cylinder, of approximate dimensions  2.73 mm (height) $\times$ 3.9 mm (dia.), enclosed in a OFHC Cu cylindrical enclosure. The dielectric rod is placed onto the sample, which in turn is partially masked by a thin metal mask with a circular hole in order to maintain the azimuthal symmetry, and it is in close contact with the bottom wall of the enclosure. The high permittivity of rutile ($\epsilon_r \sim$100, for a detailed characterization see \cite{KleinJAP95}) ensures that the evanescent electromagnetic field drops quickly to zero out of the dielectric. The DR works in transmission with very weak coupling, thus making corrections of the directly measured quality factor $Q$ unnecessary \cite{collin}. The measurements here reported are all in the linear regime, as directly checked by us by varying the incident power over a factor of 10 without observing any variation in the response.  A sketch of the resonator is reported in Fig.\ref{figresonator}a. The resonator was placed in a He-flow superconducting cryomagnet, and magnetic fields up and over the upper critical field $H_{c2}$ could be applied perpendicular to the sample plane. At fixed temperatures (stability $<0.001\;$K) the field was slowly swept. At each magnetic field the resonance curve was measured with a Vector Network Analyzer and fitted to a Lorentzian, yielding $Q$ and the resonance frequency $\nu_0$.

All samples were thin enough to ensure the realization of the thin-film approximation in the microwave measurements \cite{silvaSUST96}. Thus, the field variation of the complex resistivity $\Delta\rho(H)=\rho(H)-\rho(0)=\Delta\rho_1 + \mathrm{i}\Delta\rho_2$ was obtained from $Q$ and $\nu_0$ as:
\begin{equation}
\label{eq:rhoexp}
   \frac{ \Delta\rho(H)}{G d }=\left[ \frac{1}{Q(H)}-\frac{1}{Q(0)}  -2\mathrm{i} \left( \frac{\nu_0(H)-\nu_0(0)}{\nu_0(0)} \right) \right]
\end{equation}
where $G$ is a geometrical factor and $d=2d_{Nb}+d_F$ is the total thickness of the sample. The reduced thickness implies that screening is vanishing small approaching $T_c$, so that data close to the normal state cannot be measured with the present technique. We note that $G$ and $d$ act as mere scaling factors. In Figure \ref{figresonator} we report the evolution of the resonance curve upon sweeping the magnetic field. $Q$ and $\nu_0$ change with the field (the resonance curve broadens and the peak shifts), reflecting changes in $\rho_1$ and $\rho_2$. From the application of Eq.s (\ref{eq:rhovm}), (\ref{eq:rhoexp}) and the procedure described in \cite{pompeoPRB08} we found that, to a good approximation, $\rho_{ff}\simeq\Delta\rho_1$ (being $\Delta\rho_1$ a lower limit for $\rho_{ff}$ when the correction could be detected). For the purposes of this paper, we could safely take $\rho_{ff}\simeq\Delta\rho_1$.

In some samples and in a Nb sample (for reference) we checked at several temperatures that the microwave resistivity as measured with the DR and the Corbino disk yielded the same results.

\begin{figure}[h]
\centerline{\includegraphics[width=7.cm]{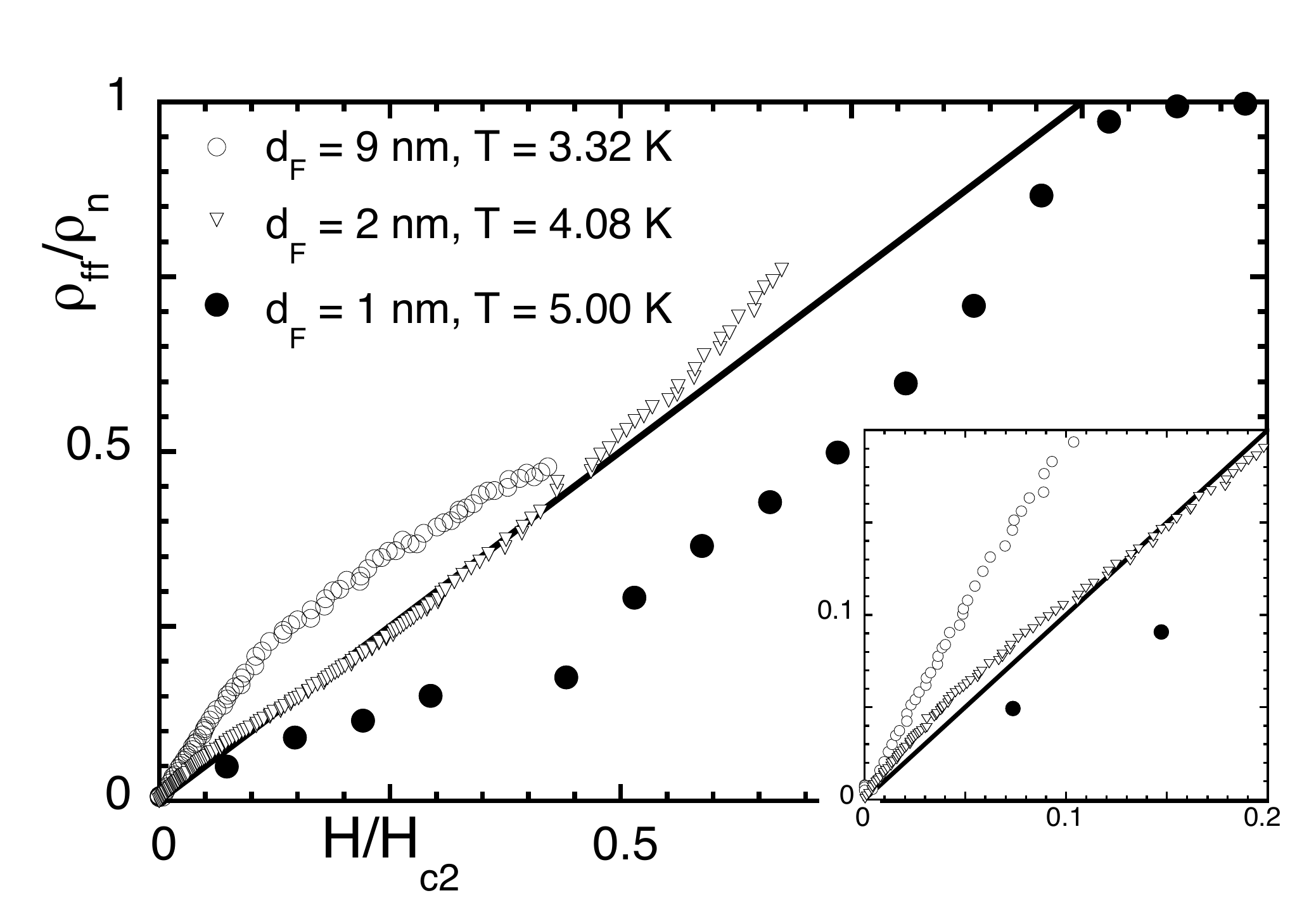}}
  \caption{Normalized flux flow resistivity vs. reduced field, $\rho_{ff}/\rho_n$ vs. $H/H_{c2}$, at approximately the same reduced temperature $T/T_c\approx 0.8$ for three trilayers with $d_F= 1, 2, 9$nm. Continuous line: BS prediction, Eq.(\ref{eq:rhoBS}), with $c=1$. None of the samples follows the BS model. Inset: enlargement at small reduced fields.}
\label{figrhoff}
\end{figure}
\section{Results and discussion}
\label{results}
In Fig.\ref{figrhoff} we report the normalized flux flow resistivity in the three samples studied at similar reduced temperature $T/T_c\approx 0.8$, as a function of the reduced magnetic field. The sample with $d_F= $1 nm has been measured with the Corbino disk. In this plot in reduced units one would expect a similar, if not the same, behavior for $\rho_{ff}$. However, despite the same reduced temperature, the three samples behave dramatically different. It is apparent that, with increasing $d_F$, the field dependence of $\rho_{ff}$ progressively changes curvature, from upward ($d_F=$1 nm) to downward ($d_F=9$nm). Even more striking, the large difference between the samples with $d_F=$ 1 and 2 nm: a close analysis (see the enlargement in the inset of Fig. \ref{figrhoff}) shows that even at low fields there is no proportionality $\rho_{ff}\propto H$, thus invalidating the BS model in all the samples.

\begin{figure}[h]
\centerline{\includegraphics[width=8.5cm]{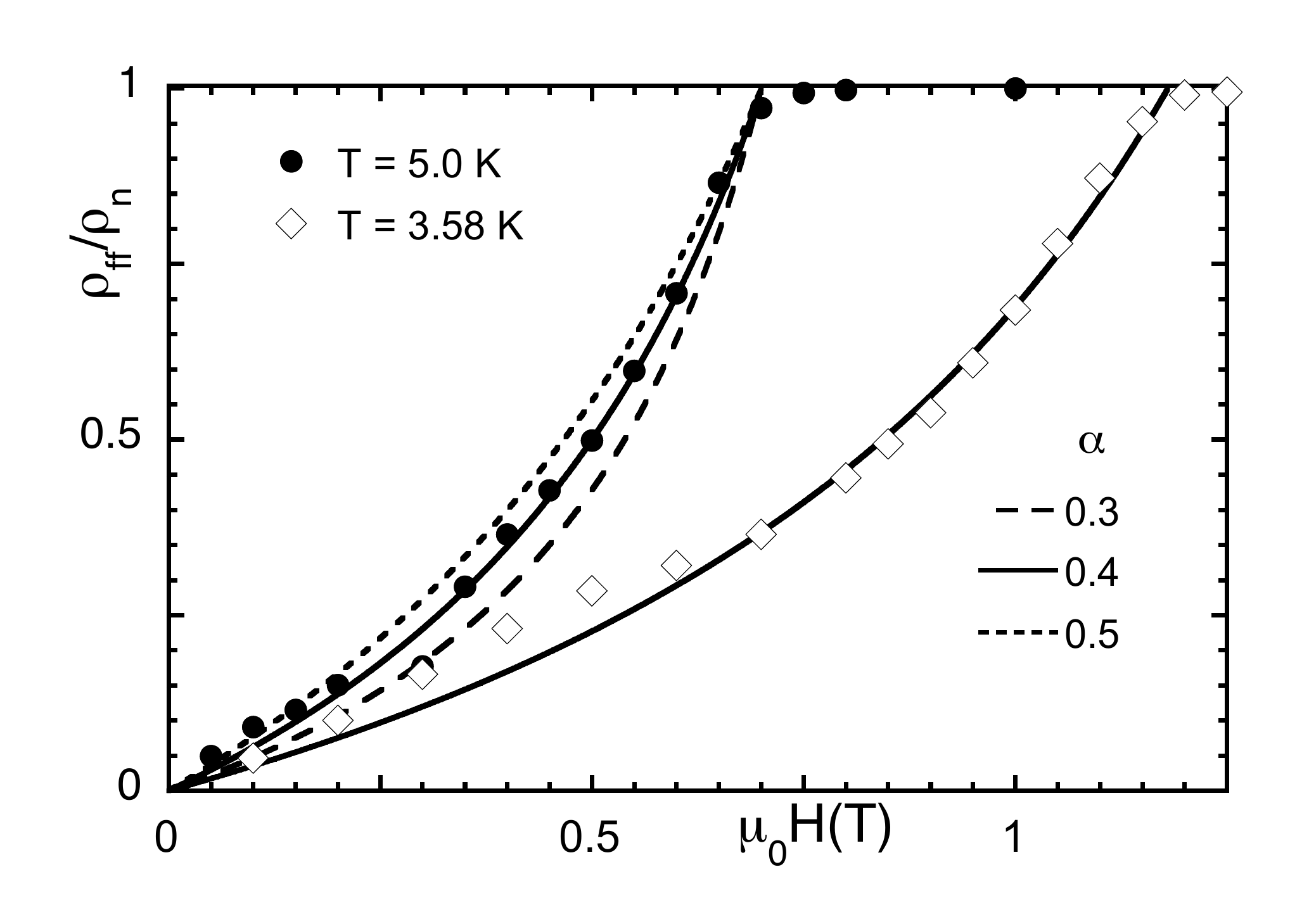}}
  \caption{Normalized flux flow resistivity vs. applied field, $\rho_{ff}/\rho_n$ vs. $H$, in the sample with $d_F=$1 nm, at 5 K (full dots, same data as in Fig.\ref{figrhoff}) and 3.58 K (open diamonds). Continuous line: TDGL calculation, Eq.(\ref{eq:rhoffGL}), with $\alpha=0.4$. By varying only the value of $H_{c2}$ the TDGL calculation fits also data at a different temperature. Short-dash and long-dash lines exemplify the sensitivity of the TDGL calculation to the parameter $\alpha$.}
\label{figrhoffGL}
\end{figure}

We discuss first the sample at $d_F=$1 nm. First of all, it should be made clear that the upward curvature, and the fact that $\rho_{ff} < H/H_{c2}$ cannot be ascribed to pinning or creep: the present measurements have been taken with the Corbino disk, and the full frequency dependence of the real part of the resistivity has been recorded (typical frequency sweeps, and fits by Eq.(\ref{eq:rhovm}), have been reported in \cite{silvaSUST11}). The upward curvature reflects an overestimate of $\rho_{ff}$ by the BS model. We then compared the results to more refined theories for the FF. We compared the data with the microscopic theory by Larkin and Ovchinnikov (LO) \cite{LO}, but were unable to obtain more than qualitative agreement in the low field and $H\rightarrow H_{c2}$ limit. However, we found that a full calculation within the time-dependent Ginzburg-Landau framework (TDGL) \cite{troyPRB93} matched very well our data. The calculation yielded, for the normalized flux flow resistivity:
\begin{equation}
\label{eq:rhoffGL}
   \frac{\rho_{ff}}{\rho_n}=\frac{1}{1+\left( \mu_0 H_{c2}-B\right)/\alpha B}
\end{equation}
where $\alpha\approx 0.4$, as calculated in \cite{liangPRB2010}. The BS model is recovered for $\alpha=1$.  In Fig.\ref{figrhoffGL} we report the same data for the sample with $d_F=$1 nm as in Fig.\ref{figrhoff}, together with the curves calculated on the basis of Eq.(\ref{eq:rhoffGL}) for $\alpha=$0.3, 0.4, 0.5. To check further the agreement between the data and the TDGL expression, we measured $\rho_{ff}$ at a different temperature. As can be seen, with the simple change of the value of $H_{c2}$, Eq.(\ref{eq:rhoffGL}) fits the data at different temperature as well.
While we can state that we found agreement between our $\rho_{ff}$ data and a ``conventional" theory, so that at least in this respect the trilayer with very short $d_F$ behaves essentially like an ordinary superconductor, the same is not true for the trilayers with larger $d_F$. We argue that the increasing $d_F$ induces a progressive crossover between a ``conventional" behavior and a novel flux flow regime. The data in the sample with $d_F=$9 nm exceed by far the BS expectation, indicating that \textit{additional} losses are induced by the magnetic field. The emerging of this feature can be observed at low fields in the sample with $d_F=$2 nm (see inset of Fig.\ref{figrhoff}). While we have not yet an explanation of the phenomenon, we can speculate over a possible scenario. At very small $d_F$ the F layer acts simply to reduce $T_c$ (with respect to a Nb sample 30 nm thick, where $T_c\simeq$7.5 K \cite{pompeoPhC10}) but leaving all the main superconducting features unaffected, including the spatial uniformity of the order parameter. With increasing $d_F$, the order parameter acquires a modulation across the trilayer (connected to the observed oscillations of $T_c$ with $d_F$  \cite{jiangPRL95,kontosPRL02,zdravkovPRL06}). The magnetic field may trigger this delicate state, inducing a relatively wide region where the order parameter is substantially depressed. The electromagnetic field, uniform along the thickness of thin films, would give a response as if part of the trilayer would be essentially normal (effective thickness smaller than the geometrical thickness), and  dissipation larger than expected would be observed. While this is at present only a speculation, it indicates that an extension of the measurements in samples with different $d_F$, and possibly with different F materials, would be useful to clarify the anomalous flux-flow behavior here observed.
\section{Summary}
\label{conc}
We have studied the flux flow resistivity of Nb/PdNi/Nb trilayers. We have found a conventional behavior of $\rho_{ff}$, different from the Bardeem-Stephen model but well described by a TDGL calculation, when the ferromagnetic thickness is extremely small. When the thickness of the ferromagnetic layer increases, $\rho_{ff}$ exhibits a previously unreported field dependence. A peculiar feature is that it exceeds the Bardeen-Stephen value, $\rho_n H/H_{c2}$, in almost the entire field range. A quantitative explanation of the phenomenon is not given at present. 
\\
\\
This work has been partially supported by an Italian MIUR-PRIN 2007 project.




\begin{thebibliography}{00}

%
\bibitem{blatterRMP94} G. Blatter, M. V. Feigel'man, V. B. Geshkenbein, A. I. Larkin and V. M. Vinokur, Rev. Mod. Phys. 66 (1994) 1125.
%
\bibitem{tomaschPRB88} W. J. Tomasch, H. A. Blackstead, S. T. Ruggiero, P. J. McGinn, J. R. Clem, K. Shen, J. W. Weber and D. Boyne, Phys. Rev. B 37 (1988) 9864.
%
\bibitem{GR} J. Gittleman and B. Rosenblum, Phys. Rev. Lett. 16 (1966) 734.
%
\bibitem{CC} M.W. Coffey and J.R. Clem, Phys. Rev. Lett. {67} (1991) 386.
%
\bibitem{brandtPRL91} E. H. Brandt, {Phys. Rev. Lett.} {67} (1991) 2219.
%
\bibitem{MStheory} T. Hocquet, P. Mathieu, and Y. Simon, Phys. Rev. B {46} (1992) 1061; B. Placais, P. Mathieu, Y. Simon, E. B. Sonin, and K. B. Traito, Phys. Rev. B {54} (1996) 13083.
%
\bibitem{pompeoPRB08} N. Pompeo and E. Silva, Phys. Rev. B {78} (2008) 094503.
%
\bibitem{cirilloSUST11} C. Cirillo, E. A. Ilyina and C. Attanasio, Supercond. Sci. Technol.  {24} (2011) 024017.
%
\bibitem{pompeoSUST05} N. Pompeo, R. Marcon, L. M\'echin and E. Silva, Supercond. Sci. Technol. {18} (2005) 531.
%
\bibitem{silvaSUST11} E. Silva, N. Pompeo, S. Sarti, {Supercond. Sci. Technol.} {24} (2011) 024018.
%
\bibitem{KleinJAP95} N. Klein, C. Zuccaro, U. D\"ahne, H. Schulz, N. Tellmann, R. Kutzner, A. G. Zaitsev, and R. W\"ordenweber, J. Appl. Phys. 78 (1995) 6683.
%
\bibitem{collin} E. R. Collin,  {\it Foundation for Microwave Engineering}, 1992 McGraw-Hill International Editions.
%
\bibitem{silvaSUST96} E. Silva, M. Lanucara, R. Marcon, {Supercond. Sci. Technol.} {9} (1996) 934.
%
\bibitem{BS} J. Bardeen and M. J. Stephen, Phys. Rev. {140} (1965) A1197.
%
\bibitem{LO} A. I. Larkin and Yu. N. Ovchinnikov,  in {\it Nonequilibrium Superconductivity}, ed. by D. N. Langenberg and A. I. Larkin, 1986, Elsevier, Amsterdam.
%
\bibitem{troyPRB93} R. J. Troy and A. T. Dorsey, Phys. Rev. B  {47} (1993) 2715, and references therein.
%
\bibitem{liangPRB2010} M. Liang, M. N. Kunchur, J. Hua, Z. Xhiao, Phys. Rev. B  {82} (2010) 064502.
%
\bibitem{pompeoPhC10} N. Pompeo, E. Silva, S. Sarti, C. Attanasio, C. Cirillo, {Physica C} {470} (2010) 901.
%
\bibitem{jiangPRL95}  J. S. Jiang, D. Davidovic, H. Reich Daniel and C. L. Chien, Phys. Rev. Lett. {74} (1995) 314.
%
\bibitem{kontosPRL02}  T. Kontos, M. Aprili, J. Lesueur, F. Genet, B. Stephanidis and R. Boursier, Phys. Rev. Lett. {89} (2002) 137007.
%
\bibitem{zdravkovPRL06} V. Zdravkov, A. Sidorenko, G. Obermeier, S. Gsell, M. Schreck, C. M\"uller, S. Horn, R. Tidecks and L. R. Tagirov,  Phys. Rev. Lett. {97} (2006) 057004.


\end{thebibliography}



\end{document}